\documentclass[runningheads]{llncs}
\usepackage{graphicx}
\usepackage{amsmath}
\usepackage{amsfonts}
\usepackage{stmaryrd}
\usepackage{listings} 
\usepackage{multirow}
\usepackage{makecell}
\usepackage{xcolor}
\usepackage{etoolbox}
\usepackage{hyperref}

\lstdefinelanguage{idl}{%
  alsoletter={-?\#:},
  otherkeywords={(,),\{,\},[,]},
  morekeywords={def,fun,def-data,def-struct,match,let,error},
  morekeywords=[2]{String,Integer,Any,Boolean},
  morekeywords=[3]{(,),\{,\},[,]},
  morekeywords=[4]{\#:bar,\#:apply,\#:atomic,\#:no-defun,\#:name,\#t, \#f, \#:foo},
  sensitive=true,
  morestring=[b]",
  morecomment=[l]{;;}
}

\expandafter\patchcmd\csname \string\lstinline\endcsname{%
  \leavevmode
  \bgroup
}{%
  \leavevmode
  \ifmmode\hbox\fi
  \bgroup
}{}{%
  \typeout{Patching of \string\lstinline\space failed!}%
}

\newcommand{\LC}{\(\lambda\)-calculus}
\newcommand{\IDL}{\textit{IDL}}
\newcommand{\bb}[1]{\left\llbracket #1 \right\rrbracket}
\newcommand{\anf}[2]{\bb{#1}#2}
\newcommand{\atomic}[1]{[ #1 ]_{a}}
\newcommand{\anfSeq}[2]{\bb{#1}_{s} #2}

\newcommand{\tuple}[1]{\left< #1 \right>}
\newcommand{\VA}[0]{\mathit{VAddr}}
\newcommand{\KA}[0]{\mathit{KAddr}}

\newcommand{\cps}[2]{\bb{#1}_c #2}
\newcommand{\dir}[1]{\bb{#1}_d}
\newcommand{\trivial}{\mathit{trivial}}

\newcommand{\allAtomic}{\mathit{allA}}
\newcommand{\noneAtomic}{\mathit{noneA}}
\newcommand{\defun}{\mathit{defun}}
\newcommand{\mkBranch}{\mathit{mkBranch}}
\newcommand{\mkApply}{\mathit{mkApply}}
\newcommand{\semt}{\lstinline!semt!}
\newcommand{\go}[3]{\mathit{go}(#1 , #2 , #3)}

\mathchardef\mh="2D

\begin{document}
\title{Automating the Functional Correspondence \\
  between Higher-Order Evaluators \\ and Abstract Machines}
%
%\titlerunning{Abbreviated paper title}
% If the paper title is too long for the running head, you can set
% an abbreviated paper title here
%
\author{Maciej Buszka\inst{1}\orcidID{0000-0003-4126-1488} \and
Dariusz Biernacki\inst{1}\orcidID{0000-0002-1477-4635}}
\authorrunning{M. Buszka, D. Biernacki}
% First names are abbreviated in the running head.
% If there are more than two authors, 'et al.' is used.
%
\institute{Institute of Computer Science, University of Wrocław, Poland}
\maketitle              % typeset the header of the contribution
%
%%%%%%%%%%%%%%%%%%%%%%%%%%%%%%%%%%%%%%%%%%%%%%%%%%%%%%%%%%%%%%%%%

\begin{abstract}
  The functional correspondence is a manual derivation technique
  transforming higher-order evaluators into the semantically
  equivalent abstract machines. The transformation consists of two
  well-known program transformations: translation to
  continuation-passing style that uncovers the control flow of the
  evaluator and Reynolds's defunctionalization that generates a
  first-order transition function. Ever since the transformation was
  first described by Danvy et al. it has found numerous applications
  in connecting known evaluators and abstract machines, but also in
  discovering new abstract machines for a variety of $\lambda$-calculi
  as well as for logic-programming, imperative and object-oriented
  languages.

  We present an algorithm that automates the functional
  correspondence. The algorithm accepts an evaluator written in a
  dedicated minimal functional meta-language and it first transforms
  it to administrative normal form, which facilitates program
  analysis, before performing selective translation to
  continuation-passing style, and selective defunctionalization. The
  two selective transformations are driven by a control-flow analysis
  that is computed by an abstract interpreter obtained using the
  abstracting abstract machines methodology, which makes it possible
  to transform only the desired parts of the evaluator. The article is
  accompanied by an implementation of the algorithm in the form of a
  command-line tool that allows for automatic transformation of an
  evaluator embedded in a Racket source file and gives fine-grained
  control over the resulting machine.

  \keywords{Evaluator \and Abstract machine \and Continuation-passing
    style \and Defunctionalization.}
\end{abstract}

%%%%%%%%%%%%%%%%%%%%%%%%%%%%%%%%%%%%%%%%%%%%%%%%%%%%%%%%%%%%%%%%%

\section{Introduction}
\label{sec:introduction}

When it comes to defining or prototyping a programming language one
traditionally provides an interpreter for the language in question
(the \emph{object}-language) written in another language (the
\emph{meta}-language)~\cite{reynolds,DBLP:books/daglib/0020601}. These
definitional interpreters can be placed on a spectrum from most
abstract to most explicit. At the abstract end lie the concise
meta-circular interpreters which use meta-language constructs to
interpret the same constructs in the object-language (e.g., using
anonymous functions to model functional values, using conditionals for
\textit{if} expressions, etc.).
% Such evaluators are typically compositional and they implement the
% valuation function of a denotational
% specification~\cite{schmidt-densem86}.
In the middle one might place various evaluators with some constructs
interpreted by simpler language features (e.g., with environments
represented as lists or dictionaries instead of functions), but still
relying on the evaluation order of the meta-language. The explicit end
is occupied by first-order machine-like interpreters which use an
encoding of a stack for handling control-flow of the
object-language.
% Such evaluators typically implement abstract
% machines~\cite{functional-correspondence}.

When it comes to modelling an implementation of a programming
language, and a functional one in particular, one traditionally
constructs an abstract machine, i.e., a first-order tail-recursive
transition system for program execution. Starting with Landin's SECD
machine~\cite{landin-secd} for \LC{}, many abstract machines have been
proposed for various evaluation strategies and with differing
assumptions on capabilities of the runtime (e.g., substitution vs
environments).  Notable work includes: Krivine's
machine~\cite{krivine-machine} for call-by-name reduction, Felleisen
and Friedman's CEK machine~\cite{felleisen-cek} and Cr\'{e}gut's
machine~\cite{cregut-normal} for normalization of $\lambda$-terms in
normal order.
% Besides equipping existing calculi with an abstract machine, new
% developments also come with both higher-level operational semantics
% and a machine, e.g., in the novel field of algebraic
% effects~\cite{biernacki-algebraic-effects,hillerstrom-algebraic-effects}.
Manual construction of an abstract machine for a given evaluation
discipline can be challenging and it requires a proof of equivalence
w.r.t. the higher-level semantics, therefore methods for deriving the
machines from natural or reduction semantics have been
developed~\cite{hannan-big-step-to-am,ager-natural-semantics,refocusing,refocusing-auto,refocusing-generalized}.
% Danvy and Nielsen's refocusing framework \cite{refocusing} gave raise
% to an automatic procedure for transforming reduction semantics into an
% abstract machine \cite{refocusing-auto,refocusing-generalized}.  Ager
% shows a mechanical method of deriving abstract machines from
% L-attributed natural semantics \cite{ager-natural-semantics} while
% Hannan and Miller present derivations of abstract machines for
% call-by-value and call-by-name reduction strategies
% \cite{hannan-big-step-to-am} via program transformations.
However, one of the most fruitful and accessible abstract machine
derivation methods was developed in the realm of interpreters and
program transformations by Danvy et al. who introduced a functional
correspondence between higher-order evaluators and abstract
machines~\cite{functional-correspondence} -- the topic of the present
work.

The functional correspondence is a realization that Reynolds's
\cite{reynolds} transformation to continuation-passing
style~\footnote{The transformation used by Reynolds was later
  formalized by Plotkin as call-by-value CPS
  translation~\cite{DBLP:journals/tcs/Plotkin75}.}  and
defunctionalization, which allow one to transform higher-order,
meta-circular, compositional definitional interpreters into
first-order, tail-recursive ones, can be seen as a general method of
actually transforming an encoding of a denotational or natural
semantics into an encoding of an equivalent abstract machine. The
technique has proven to be indispensable for deriving a
correct-by-construction abstract machine given an evaluator in a
diverse set of languages and calculi including normal and applicative
order \LC{} evaluation \cite{functional-correspondence} and
normalization~\cite{DBLP:conf/aplas/BiernackaBCD20}, call-by-need
strategy~\cite{ager-call-by-need} and \textit{Haskell}'s STG language
\cite{pirog-stg}, logic engine \cite{biernacki-logic-engine},
delimited control~\cite{biernacka-delimited-continuations},
computational effects~\cite{ager-monadic-evaluators}, object-oriented
calculi~\cite{danvy-object-oriented} and \textit{Coq}'s tactic
language~\cite{jedynak-ltac}. Despite these successes and its
mechanical nature, the functional correspondence has not yet been
transformed into a working tool which would perform the derivation
automatically.

The goal of this work is to give an algorithmic presentation of the
functional correspondence that has been implemented by the first
author as a semantics transformer. In particular, we describe the
steps required to successfully convert the human-aided derivation
method into a computer algorithm for transforming evaluators into a
representation of an abstract machine. Our approach hinges on
control-flow analysis as the basis for both selective
continuation-passing style transformation and partial
defunctionalization, and, unlike most of the works treating such
transformations~\cite{nielsen-cps,design-and-correctness-cfa}, we do
not rely on a type system. In order to obtain correct, useful and
computable analysis we employ the abstracting abstract machines
methodology (AAM) \cite{aam} which allows for deriving the analysis
from an abstract machine for the meta-language. This derivation proved
very capable in handling the non-trivial meta-language containing
records, anonymous functions and pattern matching. The resulting
analysis enables automatic transformation of user specified parts of
the interpreter as opposed to whole-program-only transformations. The
transformation, therefore, consists of: (1) transformation to
administrative normal form (ANF)~\cite{flanagan-anf} that facilitates
the subsequent steps, (2) control-flow analysis using the AAM
technique and selective (based on the analysis) CPS transformation
that makes the control flow in the evaluator explicit and idependent
from the meta-language, (3) control-flow analysis once more and
selective (again, based on the analysis) defunctionalization that
replaces selected function spaces with their first-order
representations (e.g., closures and stacks), and (4) let inlining
that cleans up after the transformation.

The algorithm has been implemented in the \emph{Haskell} programming
language giving raise to a tool --- \texttt{semt} --- performing the
transformation. The tool accepts evaluators embedded in Racket source
files. Full Racket language is available for testing the evaluators.
We tested the tool on multiple interpreters for a
diverse set of programming language calculi. It is available at:
\begin{center}
  \url{https://github.com/mbuszka/semantic-transformer}
\end{center}

The rest of this article is structured as follows: In
Section~\ref{sec:idl}, we introduce the \textit{Interpreter Definition
  Language} which is the meta-language accepted by the transformer and
will be used in example evaluators throughout the paper. In
Section~\ref{sec:transformer}, we present the algorithmic
characterization of the functional correspondence. In
Section~\ref{sec:case-studies}, we briefly discuss the performance of
the tool on a selection of case studies. In
Section~\ref{sec:conclusions}, we point at future avenues for
improvement and conclude. In Appendix~\ref{app:funcorr}, we illustrate
the functional correspondence with a minimal example, for the readers
unfamiliar with the CPS transformation and/or defunctionalization.
Appendix~\ref{app:nbe} contains an extended example---a transformation
of a normalization-by-evaluation function for \LC{} into the
corresponding abstract machine.

% We assume that the reader is familiar with \LC{} and its semantics
% (both normal (call-by-name) and applicative (call-by-value) order
% reduction).  Familiarity with formal semantics of programming
% languages (both denotational and operational) is also assumed although
% not strictly required for understanding of the main subject of this
% article. The reader should also be experienced in using a higher-order
% functional language with pattern matching.

%%%%%%%%%%%%%%%%%%%%%%%%%%%%%%%%%%%%%%%%%%%%%%%%%%%%%%%%%%%%%%%%%

\section{Interpreters and the meta-language}
\label{sec:idl}

\begin{figure}[t]
    \centering
    \begin{lstlisting}
(def-data Term
  String
  {Abs String Term}
  {App Term Term})

(def init (x) (error "empty environment"))

(def extend (env y v)
  (fun (x) (if (eq? x y) v (env x))))

(def eval (env term)
  (match term
    ([String x] (env x))
    ({Abs x body} (fun (v) (eval (extend env x v) body)))
    ({App fn arg} ((eval env fn) (eval env arg)))))
        
(def main ([Term term]) (eval init term))
    \end{lstlisting}
    \caption{A meta-circular interpreter for \LC{}}
    \label{fig:lambda-calc-interp}
\end{figure}

\begin{figure}[t]
\begin{center}
\begingroup
\setlength{\tabcolsep}{2pt}
\begin{tabular}{rrl}
  $x, y, z, f \in \mathit{Var}$ && $r\in \mathit{StructName}$\quad$s \in \mathit{String}$ \quad $b \in \mathit{Int} \cup \mathit{Boolean} \cup \mathit{String}$\\
  $\mathit{Tp} \ni \mathit{tp} $ &::=& \lstinline!String! $\mid$ \lstinline!Integer! $\mid$ \lstinline!Boolean!\\
  $\mathit{Pattern} \ni p $ &::=& $x$ $\mid$ $b$ $\mid$ \lstinline!_! $\mid$ \lstinline!{$r$ $p\ldots$}! $\mid$ \lstinline![$\mathit{tp}$ $x$]!\\
  $\mathit{Term} \ni t$ &::=& $x$ $\mid$ $b$
              $\mid$ \lstinline!(fun ($x\ldots$) $t$)!
              $\mid$ \lstinline!($t$ $t\ldots$)!
              $\mid$ \lstinline!{$r$ $t\ldots$}!\\
              &$\mid$& \lstinline!(let $p$ $t$ $t$)!
              $\mid$ \lstinline!(match $t$ ($p$ $t$)$\ldots$)!
              $\mid$ \lstinline!(error $s$)!\\
  % $\mathit{FunDef} \ni \mathit{fd}$ &::=& \lstinline!(def $x$ ($x\ldots$) $t$)!\\
  % $\mathit{StructDef} \ni \mathit{sd}$ &::=& \lstinline!(def-struct {$r$ $x\ldots$})!\\
  % $\mathit{Field}
\end{tabular}
\endgroup
\end{center}
\caption{Abstract syntax of the \IDL{} terms}\label{fig:idl-abs-syntax}
\end{figure}

The \emph{Interpreter Definition Language} or \IDL{} is the
meta-language used by \semt{} -- a semantic transformer that we have
developed. It is a purely functional, higher-order, dynamically typed
language with strict evaluation order.  It features
named records and pattern matching which allow for convenient
modelling of abstract syntax of the object-language as well as base
types of integers, booleans and strings.  The concrete syntax is in
fully parenthesized form and the programs can be embedded in a Racket
source file using a provided library with syntax definitions.

As shown in Figure~\ref{fig:lambda-calc-interp} a typical interpreter
definition consists of several top-level function \lstinline!def!initions which may be
mutually recursive.  The \lstinline!def-data! form introduces a
datatype definition. In our case it defines a type \lstinline!Term!
for terms of \LC{}. It is a union of three types: \lstinline!String!s
representing variables of \LC{}; records with label \lstinline!Abs!
and two fields of types \lstinline!String! and \lstinline!Term!
representing abstractions; and records labeled \lstinline!App! which
contain two \lstinline!Term!s and represent applications.  A datatype
definition may refer to itself, other previously defined datatypes and
records, the base types of \lstinline!String!, \lstinline!Integer! and
\lstinline!Boolean! or a placeholder type \lstinline!Any!. The \texttt{main} function
is treated as an entry point for the evaluator and must have its
arguments annotated with their type.

The \lstinline!match! expression matches an expression against a list
of patterns.  Patterns may be variables (which will be bound to the
value being matched), wildcards \lstinline!_!, base type patterns,
e.g., \lstinline![String x]! or record patterns, such as
\lstinline!{Abs x body}!.  The \lstinline!fun! form introduces
anonymous function, \lstinline!error "..."! stops execution and
signals the error.  Finally, application of a function is written as
in Racket, i.e., as a list of expressions
(e.g., \lstinline!(eval init term)!).
The evaluator in Figure~\ref{fig:lambda-calc-interp} takes advantage
of the functional representation of environments (\lstinline!init! and
\lstinline!extend!) and it structurally recursively interprets
$\lambda$-terms (\lstinline!eval!). The evaluation strategy for the
object-language is in this case inherited from the meta-language, and,
therefore, call by value (we assumed \IDL{} strict)~\cite{reynolds}.

The abstract syntax of the \IDL{} terms is presented in Figure
\ref{fig:idl-abs-syntax}. The meta-variables $x, y, z$ denote
variables; $r$ denotes structure (aka record) names; $s$ is used to
denote string literals and $b$ is used for all literal values --
strings, integers and booleans.  The meta-variable $\mathit{tp}$ is
used in pattern matches which check whether a value is one of the
primitive types.  The patterns are referred to with variable $p$ and
may be a variable, a literal value, a wildcard, a record pattern or a
type test.  Terms are denoted with variable $t$ and are either a
variable, a literal value, an anonymous function, an application, a
record constructor, a let binding (which may destructure bound term
with a pattern), a pattern match or an error expression.

\section{Transformation}
\label{sec:transformer}

The transformation described in this section consists of three main
stages: translation to administrative normal form, selective
translation to continuation-pass\-ing style, and selective
defunctionalization. After defunctionalization the program is in the
desired form of an abstract machine.  The last step taken by the
transformer is inlining of administrative let-bindings introduced by
previous steps in order to obtain more readable results.  In the
remainder of this section we will describe the three main stages of
the transformation and the algorithm used to compute the control-flow
analysis.

\subsection{Administrative Normal Form}

\begin{figure}[t]
\begin{center}
\begingroup
\setlength{\tabcolsep}{2pt}
\begin{tabular}{rll}
  $Com \ni c $ && ::= $x$ $\mid$ $b$
  $\mid$ \lstinline!(fun ($x\ldots$) $e$)!
  $\mid$ \lstinline!($x$ $x\ldots$)!\\
  &&$\mid$ \lstinline!{$r$ $x\ldots$}!
  $\mid$ \lstinline!(match $x$ ($p$ $e$)$\ldots$)!\\
  $Anf \ni e $ && ::= $c$ 
  $\mid$ \lstinline!(let $p$ $c$ $e$)!
  $\mid$ \lstinline!(error $s$)!\\
  
  % $id\,x$ &&$= x$\\
  \hline\\
  $\bb{\cdot}$ &$\cdot$ &: $Term \times (Com \rightarrow Anf) \rightarrow Anf$\\
  $\bb{x}$ &$k$ &$= k\,x$\\
  $\bb{b}$ &$k$ &$= k\,b$\\
  $\bb{\lstinline!(fun ($x\ldots$) $\,e$)!}$ &$k$
  & $= k\, \lstinline!(fun ($x\ldots$) $\anf{e}{id}$)!$\\
  
  $\bb{\lstinline!($e_f \; e_{arg}\ldots$)!}$ &$k$ 
  &$= \anf{e_f}{\atomic{\lambda x_f . \anfSeq{e_{arg}\ldots}{\lambda (x_{arg}\ldots) . k \,\lstinline!($x_f\;x_{arg}\ldots$)!}}}$\\

  $\bb{\lstinline!(let\ $p\;e_1\;e_2$)!}$ & $k$
  &$= \anf{e_1}{\lambda c_1 . \lstinline!(let\ $p\;c_1\;\anf{e_2}{k}$)!}$\\

  $\bb{\lstinline!\{$r \; e\ldots$\}!}$ &$k$ 
  &$= \anfSeq{e\ldots}{\lambda (x\ldots) . k \,\lstinline!\{$r\;x\ldots$\}!}$\\

  $\bb{\lstinline!(match\ $e \;$($p \;e_b$))!}$ & $k$
  &$= \anf{e}{\atomic{\lambda x . k\,\lstinline!(match\ $x\;$($p\;\anf{e_b}{id}$)!}}$\\

  $\bb{\lstinline!(error\ $s$)!}$ & \_ & $= $ \lstinline!(error $s$)!\\

  \hline\\
  $[\cdot]_a$ & $\cdot$ & : $(Var \rightarrow Anf) \rightarrow Com \rightarrow Anf$\\
  $[k]_a$ & $x$ & $= k\,x$\\
  $[k]_a$ & $c$ & $= $ \lstinline!(let $x$ $c$ $(k\,x)$)!\\
  \hline\\
  $\bb{\cdot}_s$ & $\cdot$ &: $Term^* \times (Var^* \rightarrow Anf) \rightarrow Anf$\\
  $\bb{e\ldots}_s$ & $k$ & $= \go{e\ldots}{\epsilon}{k}$\\
  $\go{\epsilon}{x\ldots}{k} $ & & $= k\,(x\ldots)$\\
  $\go{e\,e_r\ldots}{x_{acc}\ldots}{k}$ & & 
  $= \anf{e}{\atomic{\lambda x . \go{e_r\ldots}{x_{acc}\ldots x}{k}}}$

\end{tabular}
\endgroup
\end{center}
\caption{ANF transformation for \IDL{}}
\label{fig:transformer-anf}
\end{figure}

The administrative normal form (ANF) \cite{flanagan-anf} is an
intermediate representation for functional languages in which all
intermediate results are let-bound to names.  This shape greatly
simplifies later transformations as programs do not have complicated
sub-expressions.  From the operational point of view, the only place
where a continuation is grown when evaluating program in ANF is a
let-binding.  This property ensures that a program in ANF is also much
easier to evaluate using an abstract machine which will be taken
advantage of in Section \ref{subsec:transformer-cfa}.  The abstract
syntax of terms in ANF and an algorithm for transforming \IDL{}
programs into such form is presented in
Figure~\ref{fig:transformer-anf}.  The terms are partitioned into
three levels: variables, commands and expressions.  Commands $c$
extend variables with values -- base literals, record constructors
(with variables as sub-terms) and abstractions (whose bodies are in
ANF); and with redexes like applications of variables and match
expressions (which match on a variable and have branches in ANF).
Expressions $e$ in ANF have the shape of a possibly empty sequence of
let-bindings ending with either an error term or a command.

The $\anf{\cdot}{\cdot}$ function, written in CPS\footnote{See Appendix A of \cite{flanagan-anf}.}, is the main
transformation function.  Its arguments are a term to be transformed
and a meta-language continuation which will be called to obtain the
term for the rest of the transformed input.  This function decomposes
the term according to the (informal) evaluation rules and uses two helper
functions.  Function $\atomic{\cdot}$ transforms a continuation
expecting a variable (which are created when transforming
commands) into one accepting any command by let-binding the passed
argument $c$ when necessary. Function $\anfSeq{\cdot}{\cdot}$
sequences computation of multiple expressions by creating a chain of
let-bindings (using $\atomic{\cdot}$) and then calling the
continuation with created variables.

\subsection{Control-Flow Analysis}
\label{subsec:transformer-cfa}

The analysis most relevant to the task of deriving abstract machines
from interpreters is the control-flow analysis.  Its objective is to
find for each expression in a program an over-approximation of a set
of functions it may evaluate to \cite{popa}.  This information can be
used in two places: when determining whether a function and
applications should be CPS transformed and for checking which
functions an expression in operator position may evaluate to.  There
are a couple of different approaches to performing this analysis
available in the literature: abstract interpretation \cite{popa},
(annotated) type systems \cite{popa} and abstract abstract
machines~\cite{aam}. We chose to employ the last approach as it allows
for derivation of the control-flow analysis from an abstract machine
for \IDL{}.  The derivation technique guarantees correctness of the
resulting interpreter and hence provides high confidence in the actual
implementation of the machine. We next present the template for
acquiring both concrete and abstract versions of the abstract machine
for \IDL{}. The former machine defines the semantics of \IDL{}; the latter
computes the CFA.
%but without details of the derivation.
% from stepping through the whole derivation. To understand
% the reasoning and insights behind the technique we refer the reader to
% the original work in~\cite{aam}.

\subsubsection{A Machine Template}

\begin{figure}[t]
\begin{center}
\begingroup
\begin{tabular}{rl}
$\nu \in \VA{}$ & $\kappa \in \KA{}\quad l \in \mathit{Label}$\quad$\sigma \in \mathit{Store}$\\

$\delta \in \mathit{PrimOp}$ & $\subseteq \mathit{Val}^* \rightarrow Val$\\

$\rho \in \mathit{Env}$ &$= \mathit{Var} \rightarrow \VA{}$\\

$\mathit{Val} \ni v$ 
& ::= $b$ $\mid$ $\delta$
    $\mid$ \lstinline!{$r\;\nu\ldots$}!
    $\mid$ $\tuple{\rho,x\ldots,e}$
    $\mid$ \lstinline!(def $x$ ($x\ldots$) $e$)!\\

$\mathit{Cont} \ni k$ & ::= $\tuple{\rho, p, e, \kappa}$ $\mid$ $\tuple{}$\\

$\mathit{PartialConf} \ni \gamma $
& ::= $\tuple{\rho, e, \kappa}_e $ $\mid$ $\tuple{\nu, \kappa}_c$\\

$\mathit{Conf} \ni \varsigma $
& ::= $\tuple{\sigma, \gamma}$\\
\end{tabular}

\begin{tabular}{|rl|}
\hline
$\tuple{\sigma, \tuple{\rho, x, \kappa}_e}$
& $\Rightarrow \tuple{\mathit{copy}_v(\rho(x), l, \sigma), \tuple{\rho(x), \kappa}_c}$\\

$\tuple{\sigma, \tuple{\rho, b^l, \kappa}_e}$
& $\Rightarrow \tuple{\sigma', \tuple{\nu, \kappa}_c}$\\
& where $\tuple{\sigma', \nu} = \mathit{alloc}_v(b, l, \sigma)$\\

$\tuple{\sigma, \tuple{\rho, \lstinline!\{$r\;x\ldots$\}!^l, \kappa}_e}$
& $\Rightarrow \tuple{\sigma', \tuple{\nu, \kappa}_c}$\\
& where $\tuple{\sigma', \nu} = \mathit{alloc}_v(\lstinline!{$r\;\rho(x)\ldots$}!, l, \sigma)$\\

$\tuple{\sigma, \tuple{\rho, \lstinline!(fun ($x\ldots$)\ $e$)!^l, \kappa}_e}$
& $\Rightarrow \tuple{\sigma', \tuple{\nu, \kappa}_c}$\\
& where $\tuple{\sigma', \nu} = \mathit{alloc}_v(\tuple{\rho, x\ldots, e}, l,\sigma)$\\

$\tuple{\sigma, \tuple{\rho, \lstinline!(let\ $p\;c^l\;e$)!, \kappa}_e}$
& $\Rightarrow \tuple{\sigma', \tuple{\rho, c, \kappa'}_e}$\\
& where $\tuple{\sigma', \kappa'} = \mathit{alloc}_k(\tuple{\rho, p, e, \kappa}, l, \sigma)$\\

$\tuple{\sigma, \tuple{\rho, \lstinline!($x\;y\ldots$)!, \kappa}_e}$
& $\Rightarrow \mathit{apply}(\sigma, \rho(x), \rho(y)\ldots, l)$\\

$\tuple{\sigma, \tuple{\rho, \lstinline!(match\ $x\;$($p\;e$)$\ldots$)!, \kappa}_e}$
& $\Rightarrow \mathit{match}(\sigma, \rho, \rho(x), \tuple{p, e}\ldots)$\\

$\tuple{\sigma, \tuple{\nu, \kappa}_c}$
& $\Rightarrow \mathit{match}(\sigma, \rho, \nu, \kappa', \tuple{p, e})$\\
& where $\tuple{\rho, p, e, \kappa'} = \mathit{deref}_k(\sigma, \kappa)$\\[2pt]

% \hline &\\[\dimexpr-\normalbaselineskip+2pt]
\hline

$ \mathit{apply}(\sigma, \nu, \nu'\ldots, \kappa, l)$
& $ = \begin{cases}
  \tuple{\sigma, \tuple{\rho[(x \mapsto \nu') \ldots], e, \kappa}_e}\\
  \quad\text{when}\;\mathit{deref}_v(\sigma, \nu) = \tuple{\rho, x\ldots, e}\\

  \tuple{\sigma, \tuple{\rho_0[(x \mapsto \nu') \ldots], e, \kappa}_e}\\
  \quad\text{when}\;\mathit{deref}_v(\sigma, \nu) = \lstinline!(def $y\;$($x\ldots$) $e$)!\\

  \tuple{\sigma', \tuple{\nu'', \kappa}_c}\\
  \quad\text{when}\;\mathit{deref}_v(\sigma, \nu) = \delta\\
  \quad\text{and}\;\tuple{\sigma', \nu''} = \mathit{alloc}_v(\delta(\sigma(\nu')\ldots), l, \sigma)
\end{cases} $ \\

$ \mathit{match}(\sigma, \rho, \nu, \kappa, \tuple{p, e}\ldots)$
& $= \tuple{\sigma, \tuple{\rho', e', \kappa}_e}$ where $\rho'$ is the environment\\
&\quad for the first matching branch with body $e'$\\
\hline
\end{tabular}
\endgroup
\end{center}
\caption{A template abstract machine for \IDL{} terms in ANF}
\label{fig:anf-abstract-machine}
\end{figure}

We will begin with a template of a machine for \IDL{} terms in
A-normal form, presented in Figure \ref{fig:anf-abstract-machine}.  It
is a CEK-style machine modified to explicitly allocate memory for
values and continuations in an abstract store.  The template is
parameterized by: implementation of the store $\sigma$ along with five
operations: $\mathit{alloc}_v$, $\mathit{alloc}_k$,
$\mathit{deref}_v$, $\mathit{deref}_k$ and $\mathit{copy}_v$;
interpretation of primitive operations $\delta$ and implementation of
$\mathit{match}$ function which interprets pattern matching.  The
store maps value addresses $\nu$ to values $v$ and continuation
addresses $\kappa$ to continuations $k$.  The environment maps program
variables to value locations.  The values on which the machine operates
are the following: base values $b$, primitive operations $\delta$,
records with addresses as fields, closures and top-level functions.
Thanks to terms being in A-normal form, there are only two kinds of
continuations which form a stack.  The stack frames
$\tuple{\rho, p, e, \kappa}$ are introduced by let-bindings. They hold
an environment $\rho$, a pattern $p$ to use for destructuring a
value, the body $e$ of a let expression and a pointer to the next
continuation $\kappa$.  The bottom of the stack is marked by the empty
continuation $\tuple{}$.  We assume that every term has a unique label
$l$ which will be used in the abstract version of the machine to implement
store addresses.

The machine configurations are pairs of a store $\sigma$ and a partial
configuration $\gamma$.  This split of configuration into two parts
will prove beneficial when we instantiate the template to
obtain an abstract interpreter.  There are two classes of partial
configurations.  An evaluation configuration contains an environment
$\rho$, an expression $e$ and a continuation pointer $\kappa$.  A
continuation configuration holds an address $\nu$ of a value that has
been computed so far and a pointer $\kappa$ to a resumption which
should be applied next.

The first case of the transition relation $\Rightarrow$ looks up a
pointer for the variable $x$ in the environment $\rho$ and switches to
continuation mode.  It modifies the store via $\mathit{copy}$
function which ensures that every occurrence of a variable has a
corresponding binding in the store.  The next three cases deal with
values by $\mathit{alloc}$ating them in the store and switching to
continuation mode.  When the machine encounters a let-binding it
allocates a continuation for the body $e$ of the expression and
proceeds to evaluate the bound command $c$ with the new pointer
$\kappa'$.  In case of applications and match expressions the
resulting configuration is decided using auxiliary functions
$\mathit{apply}$ and $\mathit{match}$, respectively.  Finally, in
continuation mode, the machine may only transition if the continuation
loaded from the address $\kappa$ is a frame.  In such a case the
machine matches the stored pattern against the value pointed-to by
$\nu$.  Otherwise $\kappa$ points to a $\tuple{}$ instead and the
machine has reached the final state.  The auxiliary function
$\mathit{apply}$ checks what kind of function is referenced by $\nu$
and proceeds accordingly.

\subsubsection{A Concrete Abstract Machine}

The machine template can now be instantiated with a store, a
$\mathit{match}$ implementation which finds the first matching branch
and interpretation for primitive operations in order to obtain a
concrete abstract machine. By choosing $\mathit{Store}$ to be a
mapping with infinite domain we can ensure that $\mathit{alloc}$ can
always return a fresh address. In this setting the store-allocated
continuations are just an implementation of a stack. The extra layer
of indirection introduced by storing values in a store can also be
disregarded as the machine operates on persistent values. Therefore,
the resulting machine, which we omit, corresponds to a CEK-style
abstract machine which is a canonical formulation for call-by-value
functional calculi~\cite{DBLP:books/daglib/0023092}.

\subsubsection{An Abstract Abstract Machine}\label{ss:aam}

\begin{figure}[t]
\begin{center}
\begingroup
\begin{tabular}{rl}
$\VA{}$ & $=\KA{}=\mathit{Label}$\\
$\widetilde{\mathit{Val}} \ni v$ 
& ::= $tp$ $\mid$ $\widetilde{\delta}$
    $\mid$ \lstinline!{$r\;\nu\ldots$}!
    $\mid$ $\tuple{\rho,x\ldots,e}$
    $\mid$ \lstinline!(def $x$ ($x\ldots$) $e$)!\\

$\sigma \in \mathit{Store} $
& $= (\VA{} \rightarrow \mathbb{P}(\widetilde{\mathit{Val}}))
  \times (\KA{} \rightarrow \mathbb{P}(\mathit{Cont}))$\\

$\mathit{alloc}_v(v, l, \tuple{\sigma_v, \sigma_k})$ 
& $= \tuple{\tuple{\sigma_v[l \mapsto \sigma_v(l)\cup\{v\}], \sigma_k}, l}$\\

$\mathit{alloc}_k(v, l, \tuple{\sigma_v, \sigma_k})$
& $= \tuple{\tuple{\sigma_v, \sigma_k[l \mapsto \sigma_k(l)\cup\{k\}]}, l}$\\

$\mathit{copy}_v(\nu, l, \tuple{\sigma_v, \sigma_k})$
& $= \tuple{\sigma_v[l \mapsto \sigma_v(l)\cup\sigma_v(\nu)], \sigma_k}$\\

$\mathit{deref}_v(l, \tuple{\sigma_v, \sigma_k})$ 
& $= \sigma_v$\\

$\tilde{\varsigma} \in \widetilde{\mathit{Conf}}$
& $ = Store\times\mathbb{P}(PartialConf)$\\
\hline
$\tuple{\sigma, C}$ 
& $\Rightarrow_a \tuple{\sigma'\sqcup\sigma, C\cup C'}$\\
& where $\sigma' = \bigsqcup\{\sigma' \mid \exists \gamma \in C. \tuple{\sigma, \gamma} \Rightarrow \tuple{\sigma', \gamma'} \}$\\
& and $C' = \{\gamma' \mid \exists \gamma \in C. \tuple{\sigma, \gamma} \Rightarrow \tuple{\sigma', \gamma'} \}$\\
\hline

\end{tabular}
\endgroup
\end{center}
\caption{An abstract abstract machine for \IDL{}}
\label{fig:aam}
\end{figure}

Let us now turn to a different instantiation of the template.  Figure
\ref{fig:aam} shows the missing pieces of an abstract abstract machine
for \IDL{}.  The abstract values use base type names $tp$ to represent
any value of that type, abstract versions of primitive operations,
records, closures and top-level functions.  The interpretation of
primitive operations must approximate their concrete counterparts.

The store is represented as a pair of finite mappings from labels to
sets of abstract values and continuations, respectively.  This
bounding of store domain and range ensures that the state-space of the
machine becomes finite and therefore can be used for computing an
analysis. To retain soundness w.r.t. the concrete abstract machine the
store must map a single address to multiple values to account for
address reuse. This style of abstraction is classical~\cite{popa} and
fairly straightforward~\cite{aam}. When instantiated with this store,
the transition relation $\Rightarrow$ becomes nondeterministic as
pointer $\mathit{deref}$erencing nondeterministically returns one of
the values available in the store.  Additionally the implementation of
the $\mathit{match}$ function is also nondeterministic in the choice
of the branch to match against.

This machine is not yet suitable for computing the analysis as the
state space is still too large since every machine configuration has
its own copy of the store.  To circumvent this problem a standard
technique of widening \cite{popa} can be employed.  In particular,
following \cite{aam}, we use a global store.  The abstract
configuration $\tilde{\varsigma}$ is a pair of a store and a set of
partial configurations.  The abstract transition $\Rightarrow_a$
performs one step of computation using $\Rightarrow$ on the global
store $\sigma$ paired with every partial configuration $\gamma$.  The
resulting stores $\sigma'$ are merged together and with the original
store to create a new, extended global store.  The partial
configurations $C'$ are added to the initial set of configurations
$C$.  The transition relation $\Rightarrow_a$ is deterministic so it
can be treated as a function.  This function is monotone on a finite
lattice and therefore is amenable to fixed-point iteration.

\subsubsection{Computing the Analysis}

With the abstract transition function in hand we can now specify the
algorithm for obtaining the analysis.  To start the abstract
interpreter we must provide it with an initial configuration: a store,
an environment, a term and a continuation pointer.  The store will be
assembled from datatype and structure definitions of the program as
well as base types.  The initial term is the body of the
\lstinline!main! function of the interpreter and the environment is
the global environment extended with \lstinline!main!'s parameters
bound to pointers to datatypes in the above-built store.  The initial
continuation is of course $\tuple{}$ and the pointer is the label of
the body of the \lstinline!main! function. The analysis is computed by performing
fixed-point iteration of $\Rightarrow_a$.  The resulting store will
contain a set of functions to which every variable (the only allowed
term) in function position may evaluate (ensured by the use of
$\mathit{copy}_v$ function).  This result will be used in Sections
\ref{subsec:selective-cps} and \ref{subsec:selective-defun}.

\begin{figure}[t]
  \centering
  \begin{tabular}{rl}
    $\cps{x}{k}$ &= \lstinline!($k$ $x$)!\\
    
    $\cps{b}{k}$ &= \lstinline!(let $x$ $b$ ($k$ $x$))!\\
    
    $\cps{\lstinline!\{$r\;x\ldots$\}!}{k}$
    &= \lstinline!(let $y$ {$r\;x\ldots$} ($k$ $y$))!\\
  
    $\cps{\lstinline!(fun #:atomic ($x\ldots$)\ $e$)!}{k}$
    &= \lstinline!(let $y$ (fun ($x\ldots$) $\dir{e}$) ($k$ $y$))!\\
  
    $\cps{\lstinline!(fun ($x\ldots$)\ $e$)!}{k}$
    &= \lstinline!(let $y$ (fun ($x\ldots k'$) $\cps{e}{k'}$) ($k$ $y$))!\\
  
    $\cps{\lstinline!($f^l\;x\ldots$)!}{k}$
    &= $ \begin{cases}
      \lstinline!($f$ $x\ldots$ $k$)! & \mathrm{when}\,\noneAtomic(l)\\
      \lstinline!(let $y$ ($f$ $x\ldots$) ($k$ $y$))! & \mathrm{when}\,\allAtomic(l)\\
    \end{cases} $\\
  
    $\cps{\lstinline!(match$\;x\;$($p\;e$)$\ldots$)!}{k}$
    &= \lstinline!(match $x$ ($p$ $\cps{e}{k}$)$\ldots$)!\\
  
    $\cps{\lstinline!(let$\;p\;c\;e$)!}{k} $
    &= $ \begin{cases}
      \lstinline!(let $p$ $\dir{c}$ $\cps{e}{k}$)! \quad \mathrm{when}\,\trivial(c)\\
      \lstinline!(let $k'$ (fun ($y$) (let $p$ $y$ $\cps{e}{k}$)) $\cps{c}{k'}$)! % &\mathrm{otherwise}
    \end{cases}$\\
  
    $\cps{\lstinline!(error$\;s$)!}{k}$ &= \lstinline!(error $s$)!
  \end{tabular}
  \caption{A translation for CPS terms}
  \label{fig:cps-cps}
\end{figure}

\subsection{Selective CPS transformation}
\label{subsec:selective-cps}
In this section we formulate an algorithm for selectively transforming
the program into continuation-passing style.  All functions (both
anonymous and top-level) marked \lstinline!#:atomic!  by the user will
be kept in direct style.  The \lstinline!main!  function is implicitly
marked as atomic since its interface should be preserved as it is an
entry point of the interpreter.  Primitive operations are treated as
atomic at call-site.  Atomic functions may call non-atomic ones by
providing the called function an identity continuation. The algorithm
uses the results of the control-flow analysis to determine atomicity of
functions to which a variable labeled $l$ in function position may
evaluate.
If all functions are atomic then $\allAtomic(l)$ holds; if none of them are
atomic then $\noneAtomic(l)$ holds. When both atomic and non-atomic functions
may be called the algorithm cannot proceed and signals an error in the
source program.

\begin{figure}[h]
  \centering
  \begin{tabular}{rl}
    $\dir{x}$ &= $x$\\
    
    $\dir{b}$ &= $b$\\
    
    $\dir{\lstinline!\{$r\;x\ldots$\}!}$
    &= \lstinline!{$r\;x\ldots$}!\\
  
    $\dir{\lstinline!(fun #:atomic ($x\ldots$)\ $e$)!}$
    &= \lstinline!(fun ($x\ldots$) $\dir{e}$)!\\
  
    $\dir{\lstinline!(fun ($x\ldots$)\ $e$)!}$
    &= \lstinline!(fun ($x\ldots k'$) $\cps{e}{k'}$)!\\
  
    $\dir{\lstinline!($f^l\;x\ldots$)!}$
    &= $ \begin{cases}
      \lstinline!($f$ $x\ldots$)! & \mathrm{when}\,\allAtomic(l)\\
      \lstinline!(let $k$ (fun ($y$) $y$) ($f$ $x\ldots$ $k$))! & \mathrm{when}\,\noneAtomic(l)\\
    \end{cases} $\\
  
    $\dir{\lstinline!(match$\;x\;$($p\;e$)$\ldots$)!}$
    &= \lstinline!(match $x$ ($p$ $\dir{e}$)$\ldots$)!\\
  
    $\dir{\lstinline!(let$\;p\;$($f^l\;y\ldots$)$\;e$)!}$
    &= \begin{lstlisting}
(let $k$ (fun ($z$) $z$)
  (let $p$ ($f$ $y\ldots$ $k$) $\dir{e}$))
    \end{lstlisting}\quad when $\noneAtomic(l)$\\
  
    $\dir{\lstinline!(let$\;p\;c\;e$)!}$
    &= \lstinline!(let $p$ $\dir{c}$ $\dir{e}$)!\\
  
    $\dir{\lstinline!(error$\;s$)!}$ &= \lstinline!(error $s$)!
  \end{tabular}
  \caption{A translation for terms which should be left in direct style}
  \label{fig:cps-direct}
\end{figure}

The algorithm consists of two mutually recursive transformations.
The first, $\cps{e}{k}$ in Figure \ref{fig:cps-cps} transforms a term $e$ into
CPS. Its second parameter is a program variable $k$ which will bind
the continuation at runtime. The second, $\dir{e}$ in Figure \ref{fig:cps-direct}
transforms a term $e$ which should be kept in direct style.

%   \begin{figure}[ht]
%   \centering
%   \begin{align*}
%     \trivial(x) &\quad\trivial(b)\\
%     trivial(\lstinline!(\{$r\;x\ldots$\})!)
%     &\quad\trivial(\lstinline!(fun ($x\ldots$)\ $e$)!)\\
%   %
%     \trivial(\lstinline!($f^l\;x\ldots$)!) &\iff \allAtomic(l)\\
%   %
%     \trivial(\lstinline!(match $\;x\;$($b\;e$)$\ldots$)!)%
%     &\iff \bigwedge\trivial(e)\ldots\\
%   %
%     \trivial(\lstinline!(let$\;x\;c\;e$)!)%
%     &\iff \trivial(c) \wedge \trivial(e)
%   \end{align*}
%   \caption{The $\trivial$ predicate}
%   \label{fig:cps-trivial}
% \end{figure}

The first five clauses of the CPS translation deal with values.  When
a variable is encountered it may be immediately returned by applying a
continuation.  In other cases the value must be let-bound in order to
preserve the A-normal form of the term and then the continuation is
applied to the introduced variable.  The body $e$ of an anonymous
function is translated using $\dir{e}$ when the function is marked
atomic.  When the function is not atomic a new variable $k'$ is
appended to its parameter list and its body is translated using
$\cps{e}{k'}$.  The form of an application depends on the atomicity of
functions which may be applied.  When none of them is atomic the
continuation $k$ is passed to the function.  When all of them are
atomic the result of the call is let-bound and returned by applying
the continuation $k$.  Match expression is transformed by recursing on
its branches.  Since the continuation is always a program variable no
code gets duplicated.  When transforming a let expression the
algorithm checks whether the bound command $c$ is $\trivial$ --
meaning it will call only atomic functions when evaluated
%(defined in Figure \ref{fig:cps-trivial}).
If it is, then it can remain in direct style $\dir{c}$, no new
continuation has to be introduced and the body can be transformed by
$\cps{e}{k}$.  If the command is non-trivial then a new continuation
is created and bound to $k'$.  This continuation uses a fresh variable $y$
as its parameter. Its body is the let-expression binding $y$ instead of 
command $c$ and with body $e$ transformed with the input continuation $k$.
The bound command $c$ is transformed with the newly introduced continuation $k'$.
Finally, the translation of \lstinline!error!  throws out the continuation.

The transformation for terms which should be kept in direct style
begins similarly to the CPS one -- with five clauses for values.  In
case of an application the algorithm considers two possibilities: when
all functions are atomic the call remains in direct style, when none
of them are atomic a new identity continuation $k$ is constructed and
is passed to the called function.  A match expression is again
transformed recursively.  A let binding of a call to a CPS function
gets special treatment to preserve A-normal form by chaining allocation of
identity continuation with the call.  In other cases a let binding is
transformed recursively.  An \lstinline!error! expression is left
untouched.

Each top-level function definition in a program is transformed in the
same fashion as anonymous functions. After the transformation the
program is still in ANF and can be again analyzed by the abstract
abstract machine of the previous section. CPS-transforming the
direct-style interpreter of Figure~\ref{fig:lambda-calc-interp} yields
an interpreter in CPS shown in Figure~\ref{fig:lambda-calc-interp-cps}
(after let-inlining for readability), where we assume that the
operations on environments were marked as atomic and therefore have
not changed.

\begin{figure}[t]
    \centering
\begin{lstlisting}
(def eval (env term k)
  (match term
    ([String x] (k (env x)))
    ({Abs x body}
      (k (fun (v k') (eval (extend env x v) body k'))))
    ({App fn arg}
      (eval env fn 
        (fun (fn') (eval env arg (fun (v) (fn' v k))))))))

(def main ([Term term]) (eval init term (fun (x) x)))
\end{lstlisting}
    \caption{An interpreter for \LC{} in CPS}
    \label{fig:lambda-calc-interp-cps}
\end{figure}

\subsection{Selective Defunctionalization}
\label{subsec:selective-defun}
The second step of the functional correspondence and the last stage of
the transformation is selective defunctionalization.  The goal is to
defunctionalize function spaces deemed interesting by the author of
the program.  To this end top-level and anonymous functions may be
annotated with \lstinline!#:no-defun! to skip defunctionalization of
function spaces they belong to.  In the algorithm of Figure
\ref{fig:defun} the predicate $\defun$ specifies whether a function
should be transformed.  Predicates $\mathit{primOp}$ and
$\mathit{topLevel}$ specify whether a variable refers to (taking into
account the scoping rules) primitive operation or top-level function,
respectively.  There are three cases to consider when transforming an
application.  If the variable in operator position refers to top-level
function or primitive operation it can be left as is.  Otherwise we
can utilize the results of control-flow analysis to obtain the set of
functions which may be applied.  When all of them should be
defunctionalized ($\mathit{allDefun}$) then a call to the generated
apply function is introduced, when none of them should
($\mathit{noneDefun}$) then the application is left as is.  If the
requirements are mixed then an error in the source program is
signaled.  To transform an abstraction, its free variables ($\mathit{fvs}(l)$)
are collected into a record. The apply functions are generated using $\mkApply$ as
specified in Figure \ref{fig:defun-apply} where the
$\mathit{fn}\ldots$ is a list of functions which may be applied.
After the transformation the program is no longer in A-normal form
since variables referencing top-level functions may have been
transformed into records.  However it does not pose a problem since
the majority of work has already been done and the last step --
let-inlining does not require the program to be in
ANF. Defunctionalizing the CPS interpreter of
Figure~\ref{fig:lambda-calc-interp-cps} and performing let-inlining
yields an encoding of the CEK abstract machine shown in
Figure~\ref{fig:abstract-machine-cek} (again, the environment is left
intact).

\begin{figure}[t]
\centering
\begin{tabular}{rl}
  $\bb{x}$ &= $ \begin{cases}
    \lstinline!{Prim$_x$}! & \mathrm{when}\,\mathit{primOp}(x)\\
    % x\text{is a reference to prim op}\\
    \lstinline!{Top$_x$}!  & \mathrm{when}\,\mathit{topLevel}(x) \wedge \defun(x)\\
    % x\text{is a reference to a top-level function which should be defunctionalized}\\
    x & \mathrm{otherwise}
  \end{cases} $\\

  $\bb{b}$ &= $b$\\
  
  $\bb{\lstinline!\{$r\;x\ldots$\}!}$
  &= \lstinline!{r $\bb{x}\ldots$}!\\

  $\bb{\lstinline!(fun\ ($x\ldots$)\ $e$)$^l$!}$
  &= $\begin{cases}
    \lstinline!{Fun$_l$ $\mathit{fvs}(l)$}! &\text{when }\defun(l)\\
    \lstinline!(fun ($x\ldots$) $\bb{e}$)! &\text{otherwise}
  \end{cases}$\\

  $\bb{\lstinline!(!f^{l'}\;x\ldots\lstinline!)!^l}$
  &= $\begin{cases}
    \lstinline!($f\;\bb{x}\ldots$)! &\text{when }\mathit{primOp}(f)\vee\mathit{topLevel}(f)\\
    \lstinline!(apply$_l$ $f$ $\bb{x}\ldots$)! &\text{else when }\mathit{allDefun}(l')\\
    \lstinline!($f\;\bb{x}\ldots$)! &\text{when }\mathit{noneDefun}(l')
  \end{cases}$\\

  $\bb{\lstinline!(match$\;x\;$($p\;e$)$\ldots$)!}$
  &= \lstinline!(match $x$ ($p$ $\bb{e}$)$\ldots$)!\\

  $\bb{\lstinline!(let$\;p\;c\;e$)!}$
  &= \lstinline!(let $p$ $\bb{c}$ $\bb{e}$)!\\

  $\bb{\lstinline!(error$\;s$)!}$ &= \lstinline!(error $s$)!
\end{tabular}
\caption{Selective defunctionalization algorithm for \IDL{}}
\label{fig:defun}
\end{figure}

\begin{figure}[t]
\centering
\begingroup
\setlength{\tabcolsep}{2pt}
\begin{tabular}{rrl}
  $\mkBranch(x\ldots,$&$\delta)$
  &= \lstinline!({Prim$_\delta$} ($\delta\;x\ldots$))!\\

  $\mkBranch(x\ldots,$&$\lstinline!(def $f$ ($y\ldots$) e)!)$
  &= \lstinline!({Top$_f$} ($f$ $x\ldots$))!\\

  $\mkBranch(x\ldots,$&$\lstinline!(fun ($y\ldots$) $e$)$^l$!)$
  &= \lstinline!({Fun$_l$ $\mathit{fvs}(l)$} $\bb{e}[y\mapsto x]$)!\\
\end{tabular}
\begin{tabular}{rl}
  $\mkApply(l, \mathit{fn} \ldots)$
  &= \begin{lstlisting}
(def apply$_l$ ($f$ $x\ldots$)
  (match $f$
    $\mkBranch(x\ldots, \mathit{fn})\ldots$))
  \end{lstlisting}
\end{tabular}
\endgroup
\caption{Top-level apply function generation}
\label{fig:defun-apply}
\end{figure}

\begin{figure}[t]
\begin{lstlisting}
(def-data Cont
  {Halt}
  {App1 arg env cont}
  {App2 fn cont})

(def-struct {Closure body env x})

(def eval (env term cont)
  (match term
    ([String x] (continue cont (env x)))
    ({Abs x body} (continue cont {Closure body env x}))
    ({App fn arg} (eval env fn {App1 arg env cont}))))

(def apply (fn v cont)
  (let {Fun body env x} fn)
    (eval (extend env x v) body cont))

(def continue (cont val)
  (match cont
    ({Halt} val))
    ({App1 arg env cont} (eval env arg {App2 val cont}))
    ({App2 fn cont} (apply fn val cont)))

(def main ([Term term]) (eval {Init} term {Halt}))
\end{lstlisting}
\caption{An encoding of the CEK machine for \LC{}}
\label{fig:abstract-machine-cek}
\end{figure}

%%%%%%%%%%%%%%%%%%%%%%%%%%%%%%%%%%%%%%%%%%%%%%%%%%%%%%%%%%%%%%%%%

\section{Case Studies}
\label{sec:case-studies}

We studied the efficacy of the algorithm and the implementation on
a number of programming language calculi.  Figure
\ref{fig:tested-interpreters} shows a summary of interpreters on which
we tested the transformer.  The first group of interpreters is
denotational (mostly meta-circular) in style and covers various
extensions of the base \LC{} with call-by-value evaluation order. The
additions we tested include: integers with addition, recursive
let-bindings, delimited control operators -- \textit{shift} and
\textit{reset} with CPS interpreter based
on~\cite{biernacka-delimited-continuations} and exceptions in two
styles: monadic with exceptions as values (functions return either
value or an exception) and in CPS with success and error
continuations.  The last interpreter for call-by-value in
Figure~\ref{fig:tested-interpreters} is a normalization function based
on normalization by evaluation technique transcribed
from~\cite{abel-nbe}. We find this result particularly satisfactory,
since it leads to a non-trivial and previously unpublished abstract
machine -- we give more details in Appendix~\ref{app:nbe}. The next
three interpreters correspond to big-step operational semantics for
call-by-name \LC{}, call-by-need (call-by-name with memoization) and a
simple imperative language, respectively.

Transformation of call-by-value and call-by-need \LC{} yielded
machines very similar to the CEK and Krivine machines, respectively.
We were also able to replicate the machines previously obtained via
manual application of the functional correspondence
\cite{functional-correspondence,biernacka-delimited-continuations,biernacki-logic-engine}.
The biggest differences were due to introduction of administrative
transitions in handling of applications.  This property hints at a
potential for improvement by introducing an inlining step to the
transformation.  An interesting feature of the transformation is the
ability to select which parts of the interpreter should be transformed
and which should be considered atomic. These choices are reflected in
the resulting machine, e.g., by transforming an environment look up in
call-by-need interpreter we obtain a Krivine machine which has the
search for a value in the environment embedded in its transition
rules, while marking it atomic gives us a more abstract formulation
from \cite{functional-correspondence}.  Another consequence of this
feature is that one can work with interpreters already in CPS and
essentially skip directly to defunctionalization (as tested on
micro-Prolog interpreter of \cite{biernacki-logic-engine}).

% In the remainder of this section we will present three case studies.
% The first describes an interpreter encoding the natural semantics of a
% simple imperative language and serves as a demonstration of general
% properties of the transformation and machines it produces.  The second
% example is of an interpreter for \LC{} extended with exceptions and
% exception handlers which shows how additions to the source semantics
% translate to changes in the resulting machine.

% Interestingly, the transformation can be applied to semantic
% specifications stronger than evaluators. In particular, we obtained a
% new machine performing applicative order (full) normalization of
% $\lambda$-terms from a high-level normalization function.

\begin{figure}[t]
  \begin{center}
  \begin{tabular}{c|c|c|c}
  Language & Interpreter style & Lang. Features & Result \\
  \Xhline{2\arrayrulewidth}
  \multirow{13}{*}{\makecell{call-by-value \\ \LC{}}} & denotational & $\cdot$ & CEK machine \\
  \cline{2-4}
  & denotational & integers with add & CEK with add \\
  \cline{2-4}
  & \makecell{denotational, \\ recursion via \\ environment} & \makecell{integers, recursive \\ let-bindings} & \makecell{similar to Reynolds's  \\ first-order interpreter}\\
  \cline{2-4}
  & \makecell{denotational \\ with conts.} & shift and reset & two layers of conts.\\
  \cline{2-4}
  & \makecell{denotational, \\ monadic} & \multirow{3}{*}{\makecell{exceptions \\ with handlers}} & \makecell{explicit \\ stack unwinding}\\
  \cline{2-2}\cline{4-4}
  & \makecell{denotational, \\ CPS} & & \makecell{pointer to\\ exception handler}\\
  \cline{2-4}
  & \makecell{normalization \\ by evaluation} & $\cdot$ & strong CEK machine \\
  \hline
  \makecell{call-by-name \\ \LC{}} & big-step & $\cdot$ & Krivine machine \\
  \hline
  \makecell{call-by-need \\ \LC{}} & \makecell{big-step \\ (state passing)} & memoization & lazy Krivine machine \\
  \hline
  \makecell{simple \\ imperative} & \makecell{big-step \\ (state passing)} & \makecell{conditionals, \\ while, assignment} & $\cdot$\\
  \hline
  micro-Prolog & CPS & \makecell{backtracking, \\ cut operator} & logic engine\\
  \hline
  \end{tabular}
  \end{center}
  \caption{Summary of tested interpreters}\label{fig:tested-interpreters}
  \end{figure}

%%%%%%%%%%%%%%%%%%%%%%%%%%%%%%%%%%%%%%%%%%%%%%%%%%%%%%%%%%%%%%%%%

\section{Conclusion}
\label{sec:conclusions}

% The two transformations which form the functional correspondence have
% been studied and proven correct in various settings. The
% transformation to continuation-passing style is often used in the
% context of compilation of programming languages
% \cite{appel-compiling-with-continuations}.  In particular selective
% variants of the transformation have been proposed and proven in
% context of control operators \cite{nielsen-cps}. The main approach to
% distinguishing terms which should be transformed is to provide a type
% system annotated with required information.  The defunctionalization
% can also be used as a compilation technique but it has not seen as
% much use as other transformations such as closure conversion.
% Nevertheless there are formulations of defunctionalization based on
% control-flow analysis results \cite{design-and-correctness-cfa} which
% were proven correct.  As with the CPS transformation, the information
% is also embedded in a type system. We chose to base both
% transformations on a separately computed control-flow analysis as it
% allowed me to choose the analysis which best suited the meta-language.
% By using the AAM methodology we obtained results enabling powerful
% transformations without the implementation complexity of type
% inference, requiring annotations in the source program or turning two
% type systems into constraint-based analyses.

In this article we described an algorithm, based on the functional
correspondence~\cite{functional-correspondence}, that allows for
automatic derivation of an abstract machine given an interpreter which
typically corresponds to denotational or natural semantics, allowing
the user for fine-grained control over the shape of the resulting
machine.
%One may annotate functions which should be considered atomic
% (i.e., they do not contribute to the control-flow of the machine)
%and function spaces which should be left abstract (i.e, not
%defunctionalized).
In order to enable the transformation we derived a control-flow
analysis for \IDL{} using the abstracting abstract machines
methodology. We implemented the algorithm in the \textit{Haskell}
programming language and used this tool to transform a selection of
interpreters. To the best of our knowledge this is the first, reasonably
generic, implementation of the functional correspondence.

The correctness of the tool relies on the correctness of each of the
program transformations involved in the derivation that are classic
and in some form have been proven correct in the
literature~\cite{DBLP:journals/tcs/Plotkin75,nielsen-cps,nielsen-investigation-defunctionalization,design-and-correctness-cfa},
as well as on the correctness of the control-flow analysis we take
advantage of. An extensive number of experiments we have carried out
indicates that the tool indeed is robust.

In order to improve the capabilities of \semt{} as a practical tool for
semantics engineering, the future work could include extending the set
of primitive operations and adding the ability to import arbitrary
Racket functions and provide their abstract specification.
% Another important matter is the performance (i.e., speed) of the tool.
% To this end a thorough investigation of cost and complexity of
% computing the control-flow analysis is required.
The tool could also be extended to accommodate other output formats
such as \LaTeX{} figures or low level \textit{C}
code~\cite{DBLP:books/daglib/0020601}.

Another avenue for improvement lies in extensions of the meta-language
capabilities.  Investigation of additions such as control operators,
nondeterministic choice and concurrency could yield many opportunities
for diversifying the set of interpreters (and languages) that may be
encoded in the \IDL{}.  In particular control operators could allow
for expressing the interpreter for a language with delimited control
(or algebraic
effects~\cite{biernacki-algebraic-effects,hillerstrom-algebraic-effects})
in direct style.

% Nevertheless, it appears as an interesting challenge to
% formalize a version of the functional correspondence in a proof
% assistant (e.g., Coq), with a certified abstract machine as an output.

%%%%%%%%%%%%%%%%%%%%%%%%%%%%%%%%%%%%%%%%%%%%%%%%%%%%%%%%%%%%%%%%%
%\newpage
\bibliographystyle{splncs04}
\bibliography{bibliography}

\begin{thebibliography}{10}
\providecommand{\url}[1]{\texttt{#1}}
\providecommand{\urlprefix}{URL }
\providecommand{\doi}[1]{https://doi.org/#1}

\bibitem{abel-nbe}
Abel, A.: Normalization by Evaluation: Dependent Types and Impredicativity.
  Habilitation thesis, LMU (2013)

\bibitem{ager-natural-semantics}
Ager, M.S.: From natural semantics to abstract machines. In: Etalle, S. (ed.)
  Logic Based Program Synthesis and Transformation, 14th International
  Symposium, {LOPSTR} 2004, Verona, Italy, August 26-28, 2004, Revised Selected
  Papers. Lecture Notes in Computer Science, vol.~3573, pp. 245--261. Springer
  (2004). \doi{10.1007/11506676\_16}

\bibitem{ager-interpreter-compiler}
Ager, M.S., Biernacki, D., Danvy, O., Midtgaard, J.: From interpreter to
  compiler and virtual machine: a functional derivation. BRICS Report Series
  \textbf{10}(14) (2003)

\bibitem{functional-correspondence}
Ager, M.S., Biernacki, D., Danvy, O., Midtgaard, J.: A functional
  correspondence between evaluators and abstract machines. In: Proceedings of
  the 5th International {ACM} {SIGPLAN} Conference on Principles and Practice
  of Declarative Programming, 27-29 August 2003, Uppsala, Sweden. pp. 8--19.
  {ACM} (2003). \doi{10.1145/888251.888254}

\bibitem{ager-call-by-need}
Ager, M.S., Danvy, O., Midtgaard, J.: A functional correspondence between
  call-by-need evaluators and lazy abstract machines. Inf. Process. Lett.
  \textbf{90}(5),  223--232 (2004). \doi{10.1016/j.ipl.2004.02.012}

\bibitem{ager-monadic-evaluators}
Ager, M.S., Danvy, O., Midtgaard, J.: A functional correspondence between
  monadic evaluators and abstract machines for languages with computational
  effects. Theor. Comput. Sci.  \textbf{342}(1),  149--172 (2005).
  \doi{10.1016/j.tcs.2005.06.008}

\bibitem{design-and-correctness-cfa}
Banerjee, A., Heintze, N., Riecke, J.G.: Design and correctness of program
  transformations based on control-flow analysis. In: Kobayashi, N., Pierce,
  B.C. (eds.) Theoretical Aspects of Computer Software, 4th International
  Symposium, {TACS} 2001, Sendai, Japan, October 29-31, 2001, Proceedings.
  Lecture Notes in Computer Science, vol.~2215, pp. 420--447. Springer (2001).
  \doi{10.1007/3-540-45500-0\_21}

\bibitem{DBLP:conf/aplas/BiernackaBCD20}
Biernacka, M., Biernacki, D., Charatonik, W., Drab, T.: An abstract machine for
  strong call by value. In: d.~S.~Oliveira, B.C. (ed.) Programming Languages
  and Systems - 18th Asian Symposium, {APLAS} 2020, Fukuoka, Japan, November 30
  - December 2, 2020, Proceedings. Lecture Notes in Computer Science, vol.
  12470, pp. 147--166. Springer (2020). \doi{10.1007/978-3-030-64437-6\_8}

\bibitem{biernacka-delimited-continuations}
Biernacka, M., Biernacki, D., Danvy, O.: An operational foundation for
  delimited continuations in the {CPS} hierarchy. Log. Methods Comput. Sci.
  \textbf{1}(2) (2005). \doi{10.2168/LMCS-1(2:5)2005}

\bibitem{refocusing-generalized}
Biernacka, M., Charatonik, W., Zielinska, K.: Generalized refocusing: From
  hybrid strategies to abstract machines. In: Miller, D. (ed.) 2nd
  International Conference on Formal Structures for Computation and Deduction,
  {FSCD} 2017, September 3-9, 2017, Oxford, {UK}. LIPIcs, vol.~84, pp.
  10:1--10:17. Schloss Dagstuhl - Leibniz-Zentrum f{\"{u}}r Informatik (2017).
  \doi{10.4230/LIPIcs.FSCD.2017.10}

\bibitem{biernacki-logic-engine}
Biernacki, D., Danvy, O.: From interpreter to logic engine by
  defunctionalization. In: Bruynooghe, M. (ed.) Logic Based Program Synthesis
  and Transformation, 13th International Symposium {LOPSTR} 2003, Uppsala,
  Sweden, August 25-27, 2003, Revised Selected Papers. Lecture Notes in
  Computer Science, vol.~3018, pp. 143--159. Springer (2003).
  \doi{10.1007/978-3-540-25938-1\_13}

\bibitem{biernacki-algebraic-effects}
Biernacki, D., Pir{\'{o}}g, M., Polesiuk, P., Sieczkowski, F.: Abstracting
  algebraic effects. Proc. {ACM} Program. Lang.  \textbf{3}({POPL}),  6:1--6:28
  (2019). \doi{10.1145/3290319}

\bibitem{cregut-normal}
Cr{\'{e}}gut, P.: An abstract machine for lambda-terms normalization. In:
  Proceedings of the 1990 {ACM} Conference on {LISP} and Functional
  Programming, {LFP} 1990, Nice, France, 27-29 June 1990. pp. 333--340. {ACM}
  (1990). \doi{10.1145/91556.91681}

\bibitem{danvy-object-oriented}
Danvy, O., Johannsen, J.: Inter-deriving semantic artifacts for object-oriented
  programming. J. Comput. Syst. Sci.  \textbf{76}(5),  302--323 (2010).
  \doi{10.1016/j.jcss.2009.10.004}

\bibitem{refocusing}
Danvy, O., Nielsen, L.R.: Refocusing in reduction semantics. BRICS Report
  Series  \textbf{11}(26) (2004)

\bibitem{DBLP:books/daglib/0023092}
Felleisen, M., Findler, R.B., Flatt, M.: Semantics Engineering with {PLT}
  Redex. {MIT} Press (2009)

\bibitem{felleisen-cek}
Felleisen, M., Friedman, D.P.: Control operators, the {SECD}-machine, and the
  {\(\lambda\)}-calculus. In: Wirsing, M. (ed.) Formal Description of
  Programming Concepts - {III:} Proceedings of the {IFIP} {TC} 2/WG 2.2 Working
  Conference on Formal Description of Programming Concepts - III, Ebberup,
  Denmark, 25-28 August 1986. pp. 193--222. North-Holland (1987)

\bibitem{flanagan-anf}
Flanagan, C., Sabry, A., Duba, B.F., Felleisen, M.: The essence of compiling
  with continuations. In: Cartwright, R. (ed.) Proceedings of the {ACM}
  SIGPLAN'93 Conference on Programming Language Design and Implementation
  (PLDI), Albuquerque, New Mexico, USA, June 23-25, 1993. pp. 237--247. {ACM}
  (1993). \doi{10.1145/155090.155113}

\bibitem{DBLP:books/daglib/0020601}
Friedman, D.P., Wand, M.: Essentials of programming languages {(3.} ed.). {MIT}
  Press (2008)

\bibitem{hannan-big-step-to-am}
Hannan, J., Miller, D.: From operational semantics to abstract machines.
  Mathematical Structures in Computer Science  \textbf{2}(4),  415--459 (1992).
  \doi{10.1017/S0960129500001559}

\bibitem{hillerstrom-algebraic-effects}
Hillerstr{\"{o}}m, D., Lindley, S.: Liberating effects with rows and handlers.
  In: Chapman, J., Swierstra, W. (eds.) Proceedings of the 1st International
  Workshop on Type-Driven Development, TyDe@ICFP 2016, Nara, Japan, September
  18, 2016. pp. 15--27. {ACM} (2016). \doi{10.1145/2976022.2976033}

\bibitem{aam}
Horn, D.V., Might, M.: Abstracting abstract machines. In: Hudak, P., Weirich,
  S. (eds.) Proceeding of the 15th {ACM} {SIGPLAN} international conference on
  Functional programming, {ICFP} 2010, Baltimore, Maryland, USA, September
  27-29, 2010. pp. 51--62. {ACM} (2010). \doi{10.1145/1863543.1863553}

\bibitem{jedynak-ltac}
Jedynak, W., Biernacka, M., Biernacki, D.: An operational foundation for the
  tactic language of {C}oq. In: Pe{\~{n}}a, R., Schrijvers, T. (eds.) 15th
  International Symposium on Principles and Practice of Declarative
  Programming, {PPDP} '13, Madrid, Spain, September 16-18, 2013. pp. 25--36.
  {ACM} (2013). \doi{10.1145/2505879.2505890}

\bibitem{krivine-machine}
Krivine, J.: A call-by-name lambda-calculus machine. High. Order Symb. Comput.
  \textbf{20}(3),  199--207 (2007). \doi{10.1007/s10990-007-9018-9}

\bibitem{landin-secd}
Landin, P.J.: The mechanical evaluation of expressions. Comput. J.
  \textbf{6}(4),  308--320 (1964). \doi{10.1093/comjnl/6.4.308}

\bibitem{nielsen-investigation-defunctionalization}
Nielsen, L.: A denotational investigation of defunctionalization. BRICS Report
  Series  \textbf{7} (09 2010). \doi{10.7146/brics.v7i47.20214}

\bibitem{nielsen-cps}
Nielsen, L.R.: A selective {CPS} transformation. In: Brookes, S.D., Mislove,
  M.W. (eds.) Seventeenth Conference on the Mathematical Foundations of
  Programming Semantics, {MFPS} 2001, Aarhus, Denmark, May 23-26, 2001.
  Electronic Notes in Theoretical Computer Science, vol.~45, pp. 311--331.
  Elsevier (2001). \doi{10.1016/S1571-0661(04)80969-1}

\bibitem{popa}
Nielson, F., Nielson, H.R., Hankin, C.: Principles of program analysis.
  Springer (1999). \doi{10.1007/978-3-662-03811-6}

\bibitem{pirog-stg}
Pir{\'{o}}g, M., Biernacki, D.: A systematic derivation of the {STG} machine
  verified in coq. In: Gibbons, J. (ed.) Proceedings of the 3rd {ACM} {SIGPLAN}
  Symposium on Haskell, Haskell 2010, Baltimore, MD, USA, 30 September 2010.
  pp. 25--36. {ACM} (2010). \doi{10.1145/1863523.1863528}

\bibitem{DBLP:journals/tcs/Plotkin75}
Plotkin, G.D.: Call-by-name, call-by-value and the lambda-calculus. Theor.
  Comput. Sci.  \textbf{1}(2),  125--159 (1975).
  \doi{10.1016/0304-3975(75)90017-1}

\bibitem{reynolds}
Reynolds, J.C.: Definitional interpreters for higher-order programming
  languages. High. Order Symb. Comput.  \textbf{11}(4),  363--397 (1998).
  \doi{10.1023/A:1010027404223}

\bibitem{refocusing-auto}
Sieczkowski, F., Biernacka, M., Biernacki, D.: Automating derivations of
  abstract machines from reduction semantics: - {A} generic formalization of
  refocusing in {C}oq. In: Hage, J., Moraz{\'{a}}n, M.T. (eds.) Implementation
  and Application of Functional Languages - 22nd International Symposium, {IFL}
  2010, Alphen aan den Rijn, The Netherlands, September 1-3, 2010, Revised
  Selected Papers. Lecture Notes in Computer Science, vol.~6647, pp. 72--88.
  Springer (2010). \doi{10.1007/978-3-642-24276-2\_5}

\end{thebibliography}

%%%%%%%%%%%%%%%%%%%%%%%%%%%%%%%%%%%%%%%%%%%%%%%%%%%%%%%%%%%%%%%%%
\newpage
\appendix

\section{A Primer on the Functional Correspondence}
\label{app:funcorr}

In this section we present a simple illustrative example for the
reader unfamiliar with CPS transformations and/or defunctionalization
applied in program development. We use \IDL{} as our meta-language,
and we perform the two transformations by hand, without using our tool
automating the functional correspondence.

In this example, our object-language is the built-in data type of
integers, and the interpreter we are going to transform interprets a
given natural number as its factorial (the negative integers are
arbitrarily mapped to 1):

\begin{lstlisting}
(def factorial (n)
  (match (< 0 n)
    (#t (* n (factorial (- n 1))))
    (#f 1)))

(def main ([Integer n]) (factorial n))
\end{lstlisting}
\noindent
The familiar \lstinline!factorial! function is written in direct
style, with no explicit mention of the return stack and with a nested
recursive call. The \lstinline!main! function is the entry point for
the computation.

Let us CPS transform \lstinline!factorial!. To this end, we introduce
a functional parameter \lstinline!cont! -- a continuation -- that
represents the rest of the computation before \lstinline!factorial!
returns a value to \lstinline!main!. Returning a value from the
function is expressed by passing it to the continuation
(\lstinline!(cont 1)!). The nested recursive function call becomes a
tail call by passing the function a continuation
\lstinline!(fun (var)  (cont (* n var)))!.
The initial continuation is the identity function -- once
\lstinline!factorial! completes, it returns the final result. Here is
the CPS version of the program:

\begin{lstlisting}
(def factorial (n cont)
  (match (< 0 n)
    (#t (factorial (- n 1)
                   (fun (var) (cont (* n var)))))
    (#f (cont 1))))

(def main ([Integer n]) (factorial n (fun (x) x)))
\end{lstlisting}

Next, we defunctionalize the continuations. Defunctionalization
consists in replacing a function space with a first-order data type
and a function interpreting the constructors of this data type. Each
constructor represents a function introduction in the defunctionalized
function space; the arguments of the constructor are the values of the
free variables of the corresponding function.

In our case there are two constructors, one for the continuation in
the recursive call that should remember the values of the variables
\lstinline!cont! and \lstinline!n! (call it \lstinline!Cont!), and a
0-argument one for the initial continuation (call it
\lstinline!Halt!). We then introduce the \lstinline!continue!
function that interprets the constructors accordingly and we replace
calls to continuations with calls to this function. The resulting
first-order program reads as follows:

\begin{lstlisting}
(def-struct {Cont cont n})
(def-struct {Halt })

(def continue (fn var)
  (match fn
    ({Cont cont n} (continue cont (* n var)))
    ({Halt } var)))
\end{lstlisting}

\begin{lstlisting}
(def factorial (n cont)
  (match (< 0 n)
    (#t (factorial (- n 1) {Cont cont n}))
    (#f (continue cont 1))))

(def main ([Integer n]) (factorial n {Halt}))
\end{lstlisting}

What we have obtained is a functional encoding of an abstract
machine. The machine operates in two modes: \lstinline!factorial! and
\lstinline!continue!. The mutually tail-recursive calls model machine
transitions. The data type of the defunctionalized conitnuation,
isomorphic with a list of integers, represents the stack of the
machine. The machine did not have to be invented, but instead it was
mechanically derived. This is a very simple example of the general
phenomenon known as the functional correspondence that applies to
evaluators of virtually arbitrary complexity.

\section{Normalization by Evaluation for \LC{}}
\label{app:nbe}

In this section we present a more involved case study: deriving a
previously unknown abstract machine from a normalization function for
\LC{}. The normalization function is shown in
Figure~\ref{fig:studies-nbe}. It is based on the technique called
normalization by evaluation, and this particular definition has been
adapted from~\cite{abel-nbe}. The main idea is to use standard
evaluator for call-by-value \LC{} to evaluate terms to values and then
reify them back into terms.

\begin{figure}[t!!]
  % \lstset{basicstyle=\ttfamily\color{black}\footnotesize}
\begin{lstlisting}[numbers=left,firstline=7,lastline=43]
(def-data Term
  {Var Integer}
  {App Term Term}
  {Abs Term})

(def-struct {Level Integer})
(def-struct {Fun Any})

(def cons #:atomic (val env)
  (fun #:atomic #:no-defun (n)
    (match n
      (0 val)
      (_ (env (- n 1))))))

(def reify (ceil val)
  (match val
    ({Fun f}
      {Abs (reify (+ ceil 1) (f {Level ceil}))})
    ({Level k} {Var (- ceil (+ k 1))})
    ({App f arg} {App (reify ceil f) (reify ceil arg)})))

(def apply (f arg)
  (match f
    ({Fun f} (f arg))
    (_ {App f arg})))

(def eval (expr env)
  (match expr
    ({Var n} (env n))
    ({App f arg} (apply (eval f env) (eval arg env)))
    ({Abs body} {Fun (fun #:name Closure (x) (eval body (cons x env)))})))

(def run (term)
  (reify 0 
    (eval term (fun #:atomic #:no-defun (x) (error "empty env")))))

(def main ([Term term]) (run term))
\end{lstlisting}
  \caption{A normalization function for call-by-value \LC{}}
  % \lstset{basicstyle=\ttfamily\color{black}}
  \label{fig:studies-nbe}
\end{figure}

\begin{figure}[th!!]
  % \lstset{basicstyle=\ttfamily\color{black}\footnotesize}
\begin{lstlisting}[numbers=left,firstline=29,lastline=65]
(def reify (ceil val cont)
  (match val
    ({Fun f} (apply1 f {Level ceil} {Fun1 cont (+ ceil 1)}))
    ({Level k} (continue1 cont {Var (- ceil (+ k 1))}))
    ({App f arg} (reify ceil f {App1 arg ceil cont}))))

(def apply (f arg cont1)
  (match f
    ({Fun f} (apply1 f arg cont1))
    (_ (continue cont1 {App f arg}))))

(def eval (expr env cont2)
  (match expr
    ({Var n} (continue cont2 (env n)))
    ({App f arg} (eval f env {App3 arg cont2 env}))
    ({Abs body} (continue cont2 {Fun {Closure body env}}))))

;; evaluation
(def continue (fn2 var13)
  (match fn2
    ({Fun1 cont var3} (reify var3 var13 {Fun2 cont}))
    ({App4 cont2 var12} (apply var12 var13 cont2))
    ({App3 arg cont2 env} (eval arg env {App4 cont2 var13}))
    ({Cont cont4 var16} (reify var16 var13 cont4))))

(def run (term cont4)
  (eval term (fun (x) (error "empty env")) {Cont cont4 0}))

(def apply1 (fn x cont3)
  (match fn ({Closure body env} (eval body (cons x env) cont3))))

;; reification
(def continue1 (fn1 var11)
  (match fn1
    ({Fun2 cont} (continue1 cont {Abs var11}))
    ({App2 cont var10} (continue1 cont {App var10 var11}))
    ({App1 arg ceil cont} (reify ceil arg {App2 cont var11}))
\end{lstlisting}
  \caption{A strong call-by-value machine for \LC{}}
  % \lstset{basicstyle=\ttfamily\color{black}}
  \label{fig:studies-nbe-machine}
\end{figure}

The terms use de Bruijn indices to represent bound variables.  Since
normalization requires reduction under binders the evaluator must work
with open terms.  We use de Bruijn levels (\lstinline!Level!) to model
variables in open terms.  The \lstinline!eval! function as usual
transforms a term in a given environment into a value which is
represented as a function wrapped in a \lstinline!Fun! record.  The
values also include \lstinline!Level!s and \lstinline!Term!s which are
introduced by the \lstinline!reify!  function.  The \lstinline!apply!
function handles both the standard case of applying a functional value
(case \lstinline!Fun!) and the non-standard one which occurs during
reification of the value and amounts to emitting the syntax node for
application.  The reification function (\lstinline!reify!) turns a
value back into a term.  When its argument is a \lstinline!Fun! it
applies the function \lstinline!f! to a \lstinline!Level! representing
unknown variable.  When reified, a \lstinline!Level! is turned back
into de Bruijn index.  Lastly, reification of an (syntactic)
application proceeds recursively.  The \lstinline!main! function first
evaluates a term in an empty environment and then reifies it back into
its normal form.  As usual, we keep the environment implementation
unchanged during the transformation and we annotate the functional
values to be named \lstinline!Closure!.

The transformed normalization function is presented in Figure
\ref{fig:studies-nbe-machine}.  We notice that the machine has two
classes of continuations.  The first set (handled by
\lstinline!continue1!) is responsible for the control-flow of
reification procedure.  The second set (handled by
\lstinline!continue!) is responsible for the control-flow of
evaluation and for switching to reification mode.  We observe that the
stack used by the machine consists of a prefix of only evaluation
frames and a suffix of only reification frames.  The machine switches
between evaluation and reification in three places.  In line 3
reification of a closure requires evaluation of its body therefore
machine uses \lstinline!apply1! to evaluate the closure with a
\lstinline!Level! as an argument.  The switch in other direction is
due to evaluation finishing: in line 32 a closure's body has been
evaluated and has to be reified and then enclosed in an
\lstinline!Abs! (enforced by the \lstinline!Fun2! frame); in line 35
the initial term has been reduced and the value can be reified.

The machine we obtained, to our knowledge, has not been described in
the literature. It is somewhat similar to the one mechanically
obtained by Ager et al.~\cite{ager-interpreter-compiler} who also used
the functional correspondence to derive the machine.  Their machine
uses meta-language with mutable state in order to generate fresh
identifiers for variables in open terms instead of de Bruijn levels
and it operates on compiled rather than source terms. The machine we
obtained using \semt{} is as legible as the one derived manuallly.

\end{document}